\documentclass[twocolumn,aps,prb,superscriptaddress]{revtex4}
\usepackage{ae}
\usepackage[T1]{fontenc}
\usepackage[ansinew]{inputenc}
\usepackage{amsmath}
\usepackage{amssymb}
\usepackage{graphicx}
\usepackage{color}
\usepackage[colorlinks]{hyperref}
\usepackage{epstopdf}

\begin{document}

\date{\today}

\title{Self-excited Oscillations of Charge-Spin Accumulation Due to Single-electron Tunneling}

\author{D. Radi\'{c}}
\affiliation{Department of Applied Physics, Chalmers University of Technology, SE-412
96 G{\" o}teborg, Sweden}
\affiliation{Department of Physics, University of Gothenburg, SE-412
96 G{\" o}teborg, Sweden}
\affiliation{Department of Physics,
Faculty of Science, University of Zagreb, 1001 Zagreb, Croatia}
\author{A. M. Kadigrobov}
\affiliation{Department of Physics, University of Gothenburg, SE-412
96 G{\" o}teborg, Sweden}
\affiliation{Theoretische Physik III,
Ruhr-Universit\"{a}t Bochum, D-44801 Bochum, Germany}
\author{L. Y. Gorelik}
\affiliation{Department of Applied Physics, Chalmers University of Technology, SE-412
96 G{\" o}teborg, Sweden}
\author{R. I. Shekhter}
\affiliation{Department of Physics, University of Gothenburg, SE-412
96 G{\" o}teborg, Sweden}
\author{M. Jonson}
\affiliation{Department of Physics, University of Gothenburg, SE-412
96 G{\" o}teborg, Sweden}
\affiliation{School of Engineering and
Physical Sciences, Heriot-Watt University, Edinburgh EH14 4AS,
Scotland, UK}
\affiliation{Division of Quantum Phases
and Devices, School of Physics, Konkuk University, Seoul 143-701, Korea}

\date{\today}

\begin{abstract}
We theoretically study electronic transport through a layer of quantum dots connecting two metallic leads. By the inclusion of an inductor in series with the junction, we show that steady electronic transport in such a system may be unstable with respect to temporal oscillations caused by an interplay between the Coulomb blockade of tunneling and spin accumulation in the dots. When this instability occurs, a new stable regime is reached, where the average spin and charge in the dots oscillate periodically in time. The frequency of these oscillations is typically of the order of 1GHz for realistic values of the junction parameters.
\end{abstract}

\maketitle

\section{Introduction}

During the last decade, spin-polarized electronic transport through  quantum dots (QD) connecting
two metallic ferromagnetic leads has been the  subject of  very intensive both theoretical and experimental
research with numerous applications in spin based devices such as spin valves, spin filters, spin diodes, etc.\cite{Ono,Recher,Shokri}
Ferromagnetic ordering in the leads, which causes the rates for tunneling on-to and out-from the dot to depend on the electron spin, results in the accumulation of spin in the dots \cite{Brataas,Imamura} in addition to charge. This spin accumulation affects the transport properties of the system \cite{Rudzinski,Gorelik_coulprom} and provides a way to control the spin polarization of the current by bias or gate voltages. Moreover, it was found that the Coulomb blockade (CB) phenomenon significantly affects the DC-current \cite{Helman,Schelp,Takahashi,Barnas} and the shot noise \cite{Gorelik_shot} in such systems, which has provided  new opportunities for their electrical manipulation  \cite{Hauptmann}.

In most previous papers, time-independent effects caused by the spin and charge accumulations on the dots between ferromagnetic leads have been investigated. In this paper we focus on self induced time-dependent phenomena which may arise in such systems under a static bias with an inductor included in series with the junction. We will show that under certain  conditions, time independent electronic transport across a layer of quantum dots placed between normal and magnetic leads becomes unstable, which results  in oscillations in time of spin and charge accumulations in the layer of dots. One may assign this instability to the existence of a negative differential conductance (NDC), arising due to an interplay between the Coulomb blockade and the spin blockade
phenomena. The occurrence of an NDC in systems similar to the one considered by us was reported in \cite{Weis,Rogge,Deshmukh,Elste,Souza}. However, we will show that in our case, an instability of the time-independent regime of the charge and spin flow may arise even at a positive differential
conductance of the junction.

\section{Model}
\begin{figure}
\centerline{\includegraphics[width=6.0cm]{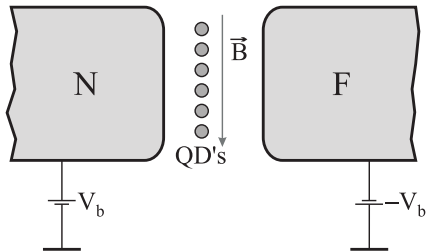}}
\caption{Charge and spin transport through a layer of quantum dots sandwiched between a normal (N) and a ferromagnetic (F) lead subjected to an external magnetic field $\vec{B}$ is studied. The junction is symmetrically biased by the voltage $2V_b$. }
\label{junction}
\end{figure}

We consider a  layer of identical quantum dots, with two spin-dependent states, connecting normal (N) and ferromagnetic (F) metallic leads. All dots in the layer are supposed to be placed at the same distance with respect to the leads (see Fig. \ref{junction}). The system under consideration is described by the Hamiltonian

\begin{equation}
\label{totHamiltonian}
H=\sum \limits_{i=N,F} H^{lead}_i + H^{QD} + \sum \limits_{i=N,F} H^{tunnel}_i,
\end{equation}
where the partial Hamiltonians

\begin{eqnarray}
\label{Hamiltonians}
H^{lead}_i &=& \sum \limits_{\vec{k},\sigma} E_{\vec{k},\sigma,i} a^{\dag}_{\vec{k},\sigma,i} a_{\vec{k},\sigma,i}, \nonumber \\
H^{QD} &=& \sum \limits_{\sigma,n} \epsilon_\sigma c^{\dag}_{\sigma,n} c_{\sigma,n} + Uc^{\dag}_{\uparrow,n} c_{\uparrow,n} c^{\dag}_{\downarrow,n} c_{\downarrow,n}, \\
H^{tunnel}_i &=& \tau_{i} \sum \limits_{\vec{k},\sigma} [a^{\dag}_{\vec{k},\sigma,i} c_{\sigma,n} + h.c.] \nonumber
\end{eqnarray}
describe electrons in the leads, in the QDs, and tunneling coupling between QDs and leads, respectively. Here $a^{\dag}_{\vec{k},\sigma,i}$ creates an electron with wave vector $\vec{k}$ and spin $\sigma=\uparrow, \downarrow$ in the corresponding lead $i=N,F$ ($i$ is the lead index); $E_{\vec{k},\sigma,N} = \varepsilon (k)$ and $E_{\vec{k},\sigma,F} = \varepsilon (k) -I_\sigma$ where $\varepsilon (k)$ is electron kinetic energy and $I_\downarrow =-I_\uparrow \equiv I$ is the ferromagnetic exchange energy; $c^{\dag}_{\sigma,n}$ creates an electron with spin $\sigma$ and energy $\epsilon_\sigma$ in the $n$-th dot ($n=1,2,...,\mathcal{N}$) counted from the identical chemical potentials of electrons in the metallic leads $\mu_{0}$; $U$ is the Coulomb interaction energy due to the double occupancy of the QD level by electrons with opposite spins. We consider the case when the inter-dot distance is much larger than the distance between the metallic leads allowing us to neglect the inter-dot Coulomb interactions. We set $\mu_{0}$ as an origin for measuring all energies. The difference between spin "up" and "down" energy levels $\Delta\epsilon=\epsilon_\uparrow-\epsilon_\downarrow$ in QD can be controlled, e.g. by the Zeeman splitting induced by the applied external magnetic field $\vec{B}$ which we will fix, for the sake of definiteness, in "down" direction ($\Delta\epsilon \approx 10^{-2} \div 10^{-1}$meV in the magnetic field $0.1 \div 1$T \cite{Schoeller}). The external magnetic field and the F-lead magnetization are taken parallel in this model, so no spin precession effects, as investigated, e.g., in Ref. \onlinecite{Konig}, are present. We will consider both the possible orientations of F-lead magnetization: along the magnetic field direction (\textbf{P}) and opposite to it (\textbf{A}). Furthermore, due to the exponential sensitivity of the tunneling matrix elements $\tau_{i}$ to the geometrical position of the QD with respect to the leads, physically interesting limiting cases can be achieved. We have studied two such cases: the N-junction, in which the QDs are closer to the N-lead ($\tau_N \gg \tau_F$), and the F-junction where they are closer to the F-lead ($\tau_F \gg \tau_N$).

\begin{figure}
\centerline{\includegraphics[width=\columnwidth]{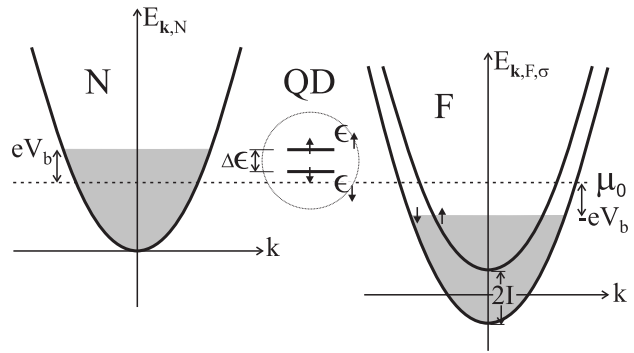}}
\caption{Electronic energy scheme (\textbf{P}-configuration) showing the nonmagnetic band (N)  and the bands for $\uparrow$ (minority) and $\downarrow$ (majority) spins in the ferromagnet (F). The latter are split by  $2I$, where  $I$ is the ferromagnetic exchange energy. The applied bias voltage $2V_b$ shifts the electron energy by $\pm eV_b$ in N/F-lead respectively from the chemical potential $\mu_{0}$. Here $\epsilon_\downarrow$ and $\epsilon_\uparrow$ are the spin-dependent energy levels in the dots.}
\label{energies}
\end{figure}

To study charge and spin transfer between leads we will consider, for the sake of simplicity, a symmetric voltage biasing of the junction (see Fig. \ref{junction}). In this case the bias voltage does not affect  the position of the QD levels $\epsilon_\sigma$, but it shifts the chemical potentials in the leads (see Fig. \ref{energies}). In this paper we will restrict our study to the case when $\epsilon_\downarrow >0$. The magnitude of the splitting $\Delta\epsilon=\epsilon_\uparrow-\epsilon_\downarrow$ is assumed to be substantially larger than the intrinsic level width and $k_B T\ll \Delta \epsilon$ ($T$ is temperature) thus providing well controlled separation of the spin states. Furthermore we will consider Coulomb blockade regime when temperature and bias voltage are much less then the double charging energy $U$. The conditions above determine the lack of charge and spin transfer if the absolute value of bias voltage $|V_{b}|$ is less then $V_\sigma=|\epsilon_\sigma /e|$ ($e$ is the electron charge). When $|V_{b}|$ exceeds $V_\downarrow$, the $\downarrow$-spin polarized tunneling arise manifesting itself as a step in the IVC (current-voltage characteristic) at $|V_{b}|=V_\downarrow$. By further increase of the bias voltage an additional step appears at $|V_{b}|=V_\uparrow$ when $\uparrow$-spin polarized tunneling is triggered. However, due to the Coulomb blockade, the presence of an electron in one state effectively blocks the current through the other one, thus affecting the IVC. We will show that a negative differential conductance can be achieved in the vicinity of $V_\uparrow$ if the tunneling rate of the electron transfer between the $\uparrow$-state in the dot and the drain electrode is low enough compared to the one for the $\downarrow$-state (spin blockade).\cite{Elste}

To analyze quantitatively kinetic properties of the system under consideration we will use rate equations for the probabilities $P_{\uparrow(\downarrow)}$ to find electron with spin $\uparrow(\downarrow)$ on the dot. They can be derived from the generalized master equation for the density matrix by adopting the Markovian approximation in the limit of weak tunneling (see Appendix).\cite{Glazman,Schoeller} Double occupation of the dots is prohibited by the Coulomb blockade thus leaving only two independent components in the rate equations (the probability to find the dot unoccupied is $P_0 =1-P_{\uparrow}-P_{\downarrow}$). The combinations $P_c = P_\uparrow + P_\downarrow $ and $P_s = P_\uparrow - P_\downarrow $ then describe charge and spin accumulation in the layer, respectively.
The rate equations describing time evolution of the average dot populations can be presented in the form

\begin{equation}
\label{sysprob}
\frac{1}{\Gamma_N}\frac{dP_\sigma}{dt} = - (1+\gamma_\sigma)P_\sigma + \left(f_N^\sigma(V)+f_F^\sigma\gamma_\sigma(V) \right)(1-P_{-\sigma}).
\end{equation}

Here $\Gamma_N=2\pi|\tau_{N}|^{2}g_{N}$, $g_{N}$ is the density of electronic states in the normal lead, which we assume to be energy independent;

\begin{equation}
\label{reltunnelrates}
\gamma_\sigma \equiv \frac{\Gamma_F^\sigma}{\Gamma_N}=   \left(\frac{\tau_{F}}{\tau_{N}}\right)^{2}\frac{g_F^\sigma}{g_N},
\end{equation}
where $g_F^\sigma$ is the spin dependent density of electronic states in the ferromagnetic lead;

\begin{eqnarray}
\label{FermiFunctions}
f_{N \sigma}(V)& = & f_{F \sigma}(-V) \equiv f\left(\epsilon_\sigma - eV \right), \nonumber \\
f(\varepsilon)&=&\frac{1}{1+e^{\varepsilon/k_B T}},
\end{eqnarray}
where $2V$ is the total voltage drop across the junction.

The average current \emph{per one dot} through the N$\rightarrow$QD junction is given by the expression

\begin{equation}
\label{current}
j(V) =e\Gamma_N \sum_{\substack{\sigma}} \left[ -P_\sigma+f_{N \sigma}(V)\left(1-P_{-\sigma}\right) \right].
\end{equation}

The geometrical asymmetry parameter $\xi \equiv |\tau_F / \tau_N|^2$ classifies 2 types of junctions: N-junction for $\xi\ll1$ and F-junction for $\xi\gg1$.\\

\section{The DC Transport}

Solving the system (\ref{sysprob}) at $V=V_{b}=const$ we obtain the time independent occupation probabilities

\begin{equation}
\label{statprob}
P^{st}_{\sigma} (V_b) =
\frac{(1+\gamma_{-\sigma})(f_{N\sigma}+f_{F\sigma }\gamma_{\sigma})-\prod\limits_{\sigma'}(f_{N \sigma' }+f_{F\sigma'}\gamma_{\sigma'})}{\prod\limits_{\sigma'}(1+\gamma_{\sigma'})-\prod\limits_{\sigma'}(f_{N\sigma'}
+f_{F\sigma'}\gamma_{\sigma'})}
\end{equation}

\begin{figure}
\centerline{\includegraphics[width=\columnwidth]{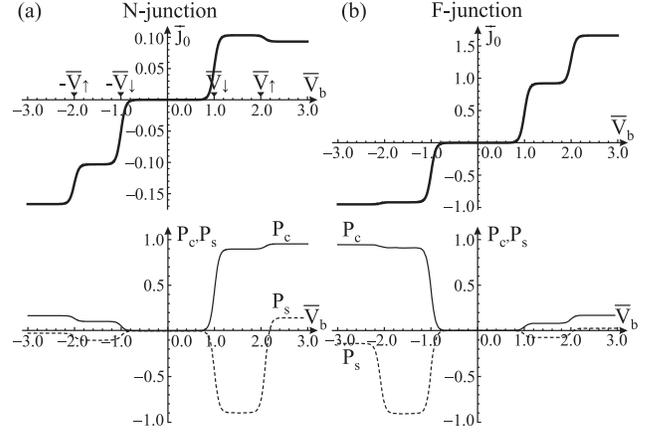}}
\caption{Single-dot IVC with corresponding on-dot charge ($P_c$) and spin ($P_s$) accumulation probabilities for: (a) an N-type junction with $\gamma_\uparrow=0.085$, $\gamma_\downarrow=0.115$ ($\xi=0.1$), and (b) an F-type junction with $\gamma_\uparrow=8.5$, $\gamma_\downarrow=11.5$ ($\xi=10$). The current is normalized as $\overline{j_0}=j_0/e\Gamma_N$. We take $\epsilon_\uparrow=2 \epsilon_\downarrow $, temperature $k_BT/\Delta\epsilon=0.05$ and introduce the dimensionless voltages $\overline{V}=eV/\Delta\epsilon$; $\overline{V}_\sigma=\epsilon_\sigma/\Delta\epsilon$. One can see that the IVC is not symmetric with respect to voltage biasing $\overline{V_b}\rightarrow-\overline{V_b}$. Also, in NF-biasing, the N-type junction produces a NDC around $\overline{V}_\uparrow=2$ in contrast to the F-type junction. For FN-biasing the NDC is absent in both cases.}
\label{IVCstat}
\end{figure}

Inserting expression (\ref{statprob}) into (\ref{current}), one gets the DC-current per dot, $j_0(V_b)$, through the junction. The IVC together with corresponding charge and spin accumulation probabilities are presented in Fig. \ref{IVCstat}.

From the expression (\ref{current}) and Fig. \ref{IVCstat} it is evident that the IVC is not symmetric with respect to a change of sign of the bias voltage. In our further investigation we will focus our attention on the phenomena occurring by activation of the spin-up channel of the electron transfer. Therefore, we will consider the  bias voltage values  $|V_b|\approx V_\uparrow$. Taking into account the fact that $|eV_b|-\epsilon_\downarrow \approx \Delta\epsilon\gg k_BT$, one can distinguish two cases:\\

a) NF-biasing ($e V_b >0$; $f_{F\sigma}(V_b)\approx 0$, $f_{N \downarrow}(V_b)\approx 1$) \\

b) FN-biasing ($e V_b< 0$; $f_{N \sigma}(V_b)\approx 0$, $f_{F \downarrow}(V_b)\approx 1$), \\

\noindent in which we obtain an approximate expression for DC-current, accurate up to the exponentially small correction $\exp(-\Delta\epsilon/k_BT)\ll 1$.\\

a) \textit{NF-biasing}

\begin{equation}
\label{statcuraprox}
j_0(V_b) \approx e\Gamma_N\gamma_\downarrow \frac{1+\gamma_\uparrow-f_{N\uparrow}(1-\gamma_\uparrow)}{(1+\gamma_{\uparrow})(1+\gamma_{\downarrow})-f_{N\uparrow}}
\end{equation}

From Eq. (\ref{statcuraprox}) it is clearly seen that the voltage dependence of the stationary IVC is entirely driven by the Fermi function $f_{N\uparrow}(V_b)$ which exhibits a step-up like behaviour at $|V_b| = V_\uparrow$. Differentiating Eq. (\ref{statcuraprox}) with respect to voltage we find the differential conductance

\begin{equation}
\label{diffcond}
\frac{dj_0}{dV_b} =- e^2\Gamma_N f'_{N\uparrow} \frac{\gamma_\downarrow (1+\gamma_\uparrow)}{\left[(1+\gamma_{\uparrow})(1+\gamma_{\downarrow})-f_{N\uparrow}\right]^2}\vartheta(\gamma_\uparrow,\gamma_\downarrow),
\end{equation}
where

\begin{eqnarray}
\label{jumpfunction}
\vartheta(\gamma_\uparrow,\gamma_\downarrow)& \equiv &\gamma_\uparrow-\gamma_\downarrow+\gamma_\uparrow \gamma_\downarrow, \nonumber \\
f'_{N\uparrow} &\equiv& \frac{df}{d\varepsilon}\Bigr|_{\epsilon_\uparrow -eV_b} < 0.
\end{eqnarray}

The corresponding current "jump" $\Delta j_0\equiv j_0(-V_\uparrow +\delta V) - j_0(-V_\uparrow- \delta V)$,
where $\Delta \epsilon \gg |e \delta V| \gg k_B T$, is

\begin{equation}
\label{currjump}
\Delta j_0=e\Gamma_N \frac{\gamma_\downarrow}{(1+\gamma_\downarrow)(\gamma_\uparrow + \gamma_\downarrow + \gamma_\uparrow \gamma_\downarrow)}\vartheta(\gamma_\uparrow,\gamma_\downarrow),
\end{equation}
in which $\vartheta(\gamma_\uparrow,\gamma_\downarrow)$ determines the sign of the "jump".\\

\begin{figure}
\centerline{\includegraphics[width=\columnwidth]{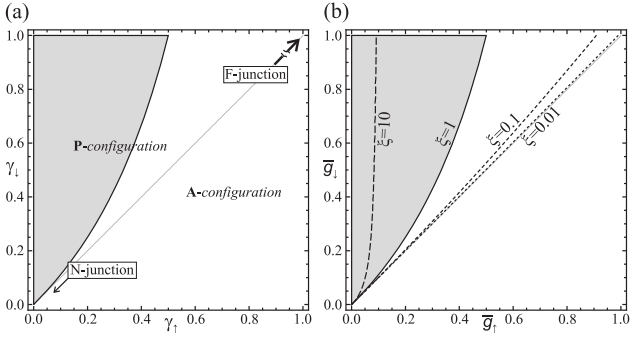}}
\caption{(a) An IVC-section with a NDC around $V_b=V_\uparrow$ appears for parameters $\gamma_\uparrow$ and $\gamma_\downarrow$ within the range corresponding to the shaded area based on the condition $\vartheta(\gamma_\uparrow,\gamma_\downarrow) <0$. (b) The same condition expressed in terms of the normalized densities of states $\overline{g}_\sigma\equiv g_F^\sigma/g_N$ for various values of the geometrical asymmetry parameter $\xi$ characterizing N-junctions ($\xi<1$) and F-junctions ($\xi>1$). The diagonal line divides the region of parameters for the \textbf{P}-configuration ($\overline{g}_\downarrow > \overline{g}_\uparrow$) and the \textbf{A}-configuration ($\overline{g}_\downarrow < \overline{g}_\uparrow$). Evidently, NDC exists in \textbf{P}-configuration only.}
\label{jumpcondpic}
\end{figure}

From Eq. (\ref{diffcond}) it can be seen that the IVC exhibits NDC, $dj_0/dV_b<0$, if $\gamma_\uparrow$ and $\gamma_\downarrow$ satisfy condition $\vartheta(\gamma_\uparrow,\gamma_\downarrow) <0$ shown graphically in Fig. \ref{jumpcondpic}. To lowest order in the small parameter formed by the ratio of the polarization parameter and the Fermi energy, $I/\epsilon_F \ll 1$, this inequality may be written as

\begin{equation}
\label{jumpcondition}
\frac{g_F^\downarrow-g_F^\uparrow}{g_N} \sim \frac{I}{\epsilon_F}>\xi.
\end{equation}

Therefore, the smaller is $\xi$, the smaller ferromagnetic polarization of F-lead is required for NDC.  Obviously,  NDC is present only in \textbf{P}-configuration. As it is seen from Eqs. (\ref{FermiFunctions}) and (\ref{diffcondR}) the differential conductance (\ref{diffcond}) is proportional to $T^{-1}$ at $V_b =-V_\uparrow$. Since it exponentially decreases at $|V_b+V_\uparrow| \gg k_B T/|e|$, the width of the "jump" in IVC is $\sim k_B T$.\\

b) \textit{FN-biasing}

\begin{equation}
\label{statcuraproxR}
j_0(V_b) \approx -e\Gamma_N \frac{f_{F\uparrow}\gamma_\uparrow(1-\gamma_\downarrow) + \gamma_\downarrow(1+\gamma_\uparrow)}{(1+\gamma_{\uparrow})(1+\gamma_{\downarrow})-f_{F\uparrow}\gamma_\uparrow\gamma_\downarrow}
\end{equation}

The differential conductance is

\begin{equation}
\label{diffcondR}
\frac{dj_0}{dV_b} = -e^2\Gamma_N f'_{F\uparrow} \frac{\gamma_\uparrow (1+\gamma_\uparrow)}{\left[(1+\gamma_{\uparrow})(1+\gamma_{\downarrow})-f_{F\uparrow}\gamma_\uparrow\gamma_\downarrow\right]^2},
\end{equation}
where

\begin{equation}
\label{FermiR}
f'_{F\uparrow} \equiv \frac{df}{d\varepsilon} \Bigr|_{\varepsilon_\uparrow+eV_b} <0.
\end{equation}
One can see that NDC does not appear in the FN-biasing of the junction.\\

The asymmetry of IVC with respect to the direction of bias voltage for junctions with magnetic leads is known and utilized e.g. for spin filtering, spin diodes etc.\cite{Recher,Shokri,Souza} In the system under consideration NDC also occurs only at the regime of an NF-biasing. Physics of this phenomenon, observed experimentally \cite{Deshmukh}, is most transparent in the case of a strongly asymmetric N-\textbf{P}-junction ($\gamma_{\uparrow}\ll\gamma_{\downarrow}\ll 1$) at low temperatures. Indeed, under such conditions and at $\epsilon_\downarrow < eV_b <\epsilon_{\uparrow}$ spin up states in the dots are not populated ($P_{\uparrow}=0$), while those for spin down are almost completely populated ($P_{\downarrow}\simeq 1$), giving the average current $j_{0}\simeq e\Gamma_{N}\gamma_{\downarrow}$. Then at $eV_{b}>\epsilon_{\uparrow}$ the up spin states start to contribute and the electrons are mostly trapped in these states ($P_{\uparrow}\simeq 1$) due to a very long escape time $\simeq (\gamma_{\uparrow}\Gamma_{N})^{-1}$ that blocks the current through the spin down states ($P_{\downarrow}\simeq \gamma_{\uparrow}/\gamma_{\downarrow} \ll 1$) by the Coulomb blockade effect. Consequently, for $eV_{b}>\epsilon_{\uparrow}$ the average current $j_{0}\simeq
e\Gamma_N(\gamma_{\uparrow}P_{\uparrow}+\gamma_{\downarrow}P_{\downarrow})\simeq 2e\Gamma_N\gamma_{\uparrow}$ is less than the average current at  $eV_{b}<\epsilon_{\uparrow}$.

\section{The Dynamical Instability of Charge-Spin Accumulations}

\begin{figure}
\centerline{\includegraphics[width=6cm]{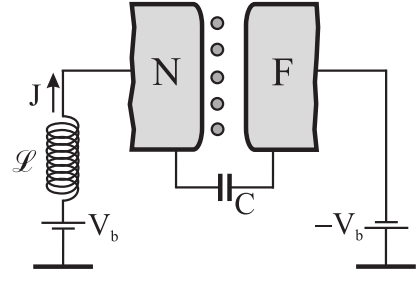}}
\caption{Circuit scheme used for the instability investigation: $C$ is the intrinsic junction capacitance, $\mathcal{L}$ is the inductance of the circuit, $J$ is the  current flowing through the inductor.}
\label{circuit}
\end{figure}

In this section we will only consider an $N$-junction ($\gamma_\sigma \leq 1$) in the regime of NF-biasing which
is the most interesting one. Really, a voltage biased circuit with a nonlinear resistor, $R(V)$, providing the section with NDC in the IVC, is usually considered as a prerequisite for dynamical instability of the system. To investigate  stability of charge/spin accumulation in our system we introduce an inductor with inductance ${\cal L}$ in series with the junction (see Fig. \ref{circuit}). We also take into account the capacitance of the junction $C \approx \mathcal{N} C_{1QD}$, where $\mathcal{N}$ is the number of quantum dots inside the junction, $C_{1QD} = \varepsilon l^2/d$ is the average capacitance per one dot ($\varepsilon$ is the dielectric constant of the layer material, $l$ is the average distance between dots, $d$ is the distance between the leads).

The tunneling and electrical processes in this system are governed by the system of four differential equations:

\begin{eqnarray}
\label{nonstatsys}
\frac{1}{\Gamma_N} \frac{dP_\sigma}{dt} + (1+\gamma_\sigma)P_\sigma - f_{N\sigma}(V) (1-P_{-\sigma}) = 0 \nonumber \\
\mathcal{L} \frac{dJ}{dt}+2V = 2V_b \\
\mathcal{N} j(V,P_\sigma) + 2C\frac{dV}{dt} = J. \nonumber
\end{eqnarray}

Here $J$ is the total current  flowing through the inductor, and the current per one dot $j(V,P_\sigma)$ is given by Eq. (\ref{current}). The set of differential equations (\ref{nonstatsys}) always has the time-independent solution

\begin{equation}
P_\sigma =P_{\sigma}^{st}, \;\; V=V_b, \;\; J=\mathcal{N} j(V_b,P_{\sigma}^{st})
\label{steady-state}
\end{equation}
with time-independent probabilities $P_{\sigma }^{st}$ given by expression (\ref{statprob}). We analyze the stability of this solution by linearizing the set of equations in terms of small deviations $\sim \exp (\lambda t)$ from the time-independent solution (\ref{steady-state}), considering the bias voltage $V_b \approx V_\uparrow$ and the temperature $k_B T \ll \Delta\epsilon$ being low enough to fulfill $f_F^\sigma(V_b)\approx 0$, $f_N^\downarrow(V_b)\approx 1$. Thus transforming system (\ref{nonstatsys}) into a set of algebraic equations, we obtain the characteristic equation of the fourth order in the Lyapunov exponent $\lambda$, i.e.

\begin{equation}
\label{characteristiceq}
a_4\lambda^4+a_3\lambda^3+a_2\lambda^2+a_1\lambda+a_0=0,
\end{equation}
with the coefficients

\begin{eqnarray}
\label{coeffs}
a_0 &=& \Gamma_N^2 A_1 \nonumber \\
a_1 &=& \Gamma_N A_2 + \mathcal{L}\mathcal{N}e\Gamma_N^3A_3\vartheta \nonumber \\
a_2 &=& 1+\mathcal{L}\mathcal{N}e\Gamma_N^2A_3 (\gamma_\uparrow+\gamma_\downarrow) + \mathcal{L}C\Gamma_N^2A_1 \nonumber \\
a_3 &=& \mathcal{L}\mathcal{N}e\Gamma_NA_3 + \mathcal{L}C\Gamma_NA_2 \nonumber \\
a_4 &=& \mathcal{L}C
\end{eqnarray}
where

\begin{eqnarray}
\label{consts}
A_1 &\equiv& (1+\gamma_\uparrow)(1+\gamma_\downarrow) - f_{N\uparrow} \nonumber \\
A_2 &\equiv& 2+\gamma_\uparrow+\gamma_\downarrow \\
A_3 &\equiv& \frac{1}{2}f_{N\uparrow}'\frac{\gamma_\downarrow(1+\gamma_\uparrow)}{(1+\gamma_\uparrow)(1+\gamma_\downarrow) - f_{N\uparrow}}. \nonumber
\end{eqnarray}

The analysis of the characteristic polynomial in the complex $\lambda$-plane, namely counting the winding number of its phase as the variable encircles the $Re(\lambda)>0$ half-plane, shows that two roots of the equation (\ref{characteristiceq}) are always real and negative, while the other two have the real part changing its sign as $\mathcal{L}$ passes through a critical value $\mathcal{L}_c$. In order to find the critical values of the inductance at which the real part of two complex conjugated Lyapunov exponents is equal to 0, we insert $\lambda= i\omega$ (were $\omega$ is real) into the characteristic equation (\ref{characteristiceq}) and obtain a set of two  equations

\begin{eqnarray}
\label{critsys}
a_4(\mathcal{L})\omega^4-a_2(\mathcal{L})\omega^2+a_0 &=& 0 \nonumber \\
a_3(\mathcal{L})\omega^2-a_1(\mathcal{L}) &=& 0
\end{eqnarray}
from which the critical inductance $\mathcal{L}_c$ and the corresponding roots $\lambda_{1,2}= \pm i \omega_c$ of the characteristic equation are found.\\

When the tunneling rates, which control the average electron populations of the dots, are the fastest rates in the system, NDC leads to the well-known electro-dynamical instability of the time-independent current flow. It is interesting to note that in the system under consideration an instability of the time-independent regime arises even in the case when the RC-time $t_{RC} \equiv |R_d|C$, where the differential resistance $R_d \equiv (\mathcal{N} d j_0/dV_b)^{-1}$ is defined by Eq. (\ref{diffcond}), is the shortest time scale in the system:

\begin{equation}
t_{RC} \ll \Gamma_N^{-1}, \;\;t_{RC} \ll \sqrt{\mathcal{L} C}.
\label{Ccondition}
\end{equation}

Taking into account realistic values of the junction parameters $l \sim 10$nm, $d \sim 1$nm, $\varepsilon \sim 10^{-11}F/m$ and the operating temperature $T \sim 50$mK, one finds that the inequalities (\ref{Ccondition}) are well fulfilled down to $\gamma_\sigma \gtrsim 10^{-2}$
and inductances $\mathcal{L}\mathcal{N}\gamma_\sigma^4 \gg 10^{-10}H$.\\

Conditions (\ref{Ccondition}) permit skipping of all terms containing $C$ in system (\ref{nonstatsys}) and coefficients (\ref{coeffs}), i.e. putting there $C=0$. In this case the set of equations (\ref{critsys}) reduces to an algebraic quadratic equation for $\mathcal{L}$ with roots

\begin{equation}
\label{critinductance}
\mathcal{L}^{(\pm)} = \frac{R_d}{\Gamma_N  } \frac{\left[ \varphi \pm \sqrt{\varphi^2-4[(1+\gamma_+)^2-1]\vartheta} \right]}{[(1+\gamma_\uparrow)(1+\gamma_\downarrow)-f_{N \uparrow}]\gamma_+},
\end{equation}
where $\vartheta(\gamma_\uparrow,\gamma_\downarrow)$ is defined by Eq. (\ref{jumpfunction}); $\gamma_+ \equiv \gamma_\uparrow+\gamma_\downarrow$ and $\varphi(\gamma_\uparrow,\gamma_\downarrow,V_b) \equiv 1-f_{N\uparrow}-2\gamma_\uparrow-\gamma_+^2$.

\begin{figure}
\centerline{\includegraphics[width=6cm]{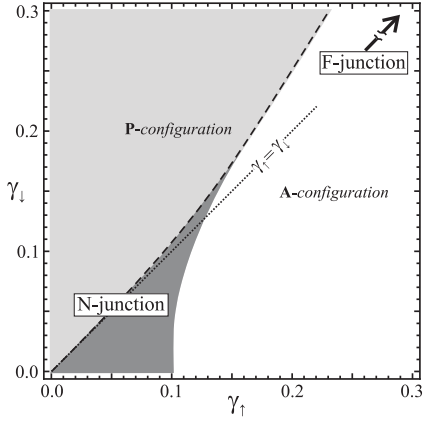}}
\caption{The instability condition in $(\gamma_\uparrow,\gamma_\downarrow)$-parameter space: the light grey area shows the range of parameters in which $R_d<0$ and the instability arises for $\mathcal{L} >\mathcal{L}_c$. The dark grey area shows the range of parameters in which $R_d > 0$ and the instability arises for $\mathcal{L}_c  < \mathcal{L} <\mathcal{L}_c^\diamond$. Here $V_b=V_\uparrow$, $k_B T/\Delta\epsilon=0.05$.}
\label{instabilitycondpic}
\end{figure}

As it follows from Eq. (\ref{critinductance}), in the case of $R_d <0$ (and hence $\vartheta <0$), one root is negative, and the other one is positive. Therefore, there is only one critical value of the inductance $\mathcal{L}_c=\mathcal{L}^{(-)}$ at which the system looses its stability. On the other hand, easily seen from coefficients (\ref{coeffs}), in the absence of external inductance ($\mathcal{L}=0$) the characteristic equation (\ref{characteristiceq}) reduces to the second order with both solutions having real parts $Re(\lambda)<0$, indicating that the fixed point is stable for any choice of the other parameters. From there it follows that in the range of parameters in which the differential resistance $R_d$ is negative (light grey area in Fig. \ref{instabilitycondpic}), the time-independent solution (\ref{steady-state}) is stable if $0  \leq \mathcal{L} \leq \mathcal{L}_c$, and, if the inductance exceeds the critical value $\mathcal{L}>\mathcal{L}_c$, the system looses its stability.\\

Following from Eq. (\ref{critinductance}), the presented system reveals one peculiar property: it may also loose its stability in the case of a positive differential resistance $R_d \geq 0$ (that is $\vartheta >0$). In the range of $\gamma_\sigma$ parameters satisfying

\begin{eqnarray}
\vartheta \geq 0; \;\;
\varphi^2 \geq 4\left[ (1+\gamma_\uparrow + \gamma_\downarrow)^2 -1 \right] \vartheta
\label{pos_dif_cond}
\end{eqnarray}
(shown as a dark grey area in Fig. \ref{instabilitycondpic}) both roots $\mathcal{L}^{(\pm)}$ are real and positive and hence there are two critical values of the inductance $\mathcal{L}_c =\mathcal{L}^{(-)}$ and $\mathcal{L}_c^\diamond =\mathcal{L}^{(+)}$. One can find that the system is unstable for inductance laying in the interval $\mathcal{L}_c  < \mathcal{L}< \mathcal{L}_c^\diamond$ and stable otherwise.

All the features mentioned above are most transparent in the limit of a highly asymmetric N-junction $\gamma_\sigma \ll 1$. In this case the critical values of the inductances $\mathcal{L}_c$ and $\mathcal{L}_c^\diamond $, as well as the corresponding frequencies $\omega_c = \omega (\mathcal{L}_c)$ and $\omega_c^\diamond  = \omega (\mathcal{L}_c^\diamond )$, are

\begin{eqnarray}
\label{critvaluesapprox}
\mathcal{L}_c \approx \frac{4}{e\Gamma_N^2 \mathcal{N} f_{N\uparrow}' \gamma_\downarrow}; \;\;
\omega_c \approx \Gamma_N \sqrt{1-f_{N\uparrow}} \nonumber \\
\mathcal{L}_c^\diamond \approx \frac{2}{e\Gamma_N^2 \mathcal{N} f_{N\uparrow}'} \frac{(1-f_{N\uparrow})^2}{\gamma_\downarrow (\gamma_\uparrow +\gamma_\downarrow)\vartheta}; \;\;
\omega_c^\diamond \approx \Gamma_N \sqrt{\vartheta}.
\end{eqnarray}

From expressions (\ref{critvaluesapprox}) it is clearly seen that in the limit $\gamma_\sigma \ll 1$ the first critical value ($\mathcal{L}_c$; $\omega_c$) does not depend on sign of differential resistance, while the second one ($\mathcal{L}_c^\diamond$; $\omega_c^\diamond$) appear as soon as IVC attains positive differential conductance ($\vartheta>0$). Therefore, the instability takes place both for $R_d< 0$ and $R_d > 0$ as soon as the inductance exceeds $\mathcal{L}_c$, but, as mentioned before, stability is established again as soon as $\mathcal{L}\geq \mathcal{L}_c^\diamond$ if the differential resistance is positive.

\begin{figure}
\centerline{\includegraphics[width=\columnwidth]{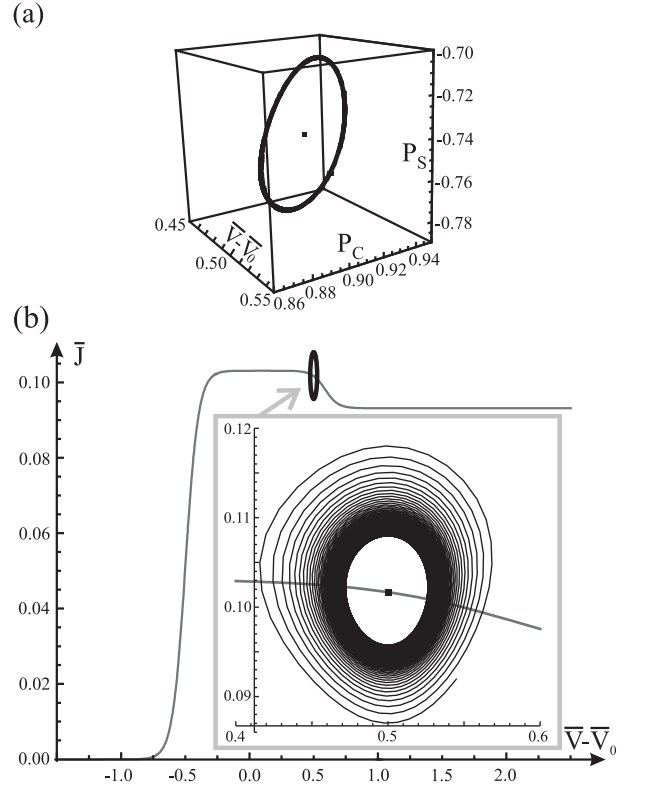}}
\caption{(a) The limit cycle in $(\overline{V},P_c,P_s)$-space around the fixed point $(0.5,0.905,-0.743)$. (b) The limit cycle in the $(\overline{V},\overline{J})$-space around the fixed point $(0.5,0.102)$ at the IVC (gray curve). We use scales: $\overline{V}=eV/\Delta\epsilon$, $\overline{J}=J/e\Gamma_N \mathcal{N}$, $\overline{\mathcal{L}} = \mathcal{L} e^2 \Gamma_N^2 \mathcal{N} / \Delta\epsilon$, $\overline{\omega}=\omega/\Gamma_N$. Voltage is measured with respect to $\overline{V}_0=(\overline{V}_\uparrow+\overline{V}_\downarrow)/2$. Here the choice of parameters is: $\overline{V}_b=\overline{V}_\uparrow$, $k_B T/\Delta\epsilon=0.05$, $\gamma_\uparrow=0.085$, $\gamma_\downarrow=0.115$, $\xi=0.1$, $\overline{\mathcal{L}}=8.0$ with critical values for instability $\overline{\mathcal{L}}_c=7.8712$, $\overline{\omega}_c=0.5457$.
This choice of parameters (\textbf{P}-configuration) leads to an IVC with NDC and the frequency of stable oscillations $\overline{\omega}=0.5456$.}
\label{limitcyclepic}
\end{figure}

\begin{figure}
\centerline{\includegraphics[width=\columnwidth]{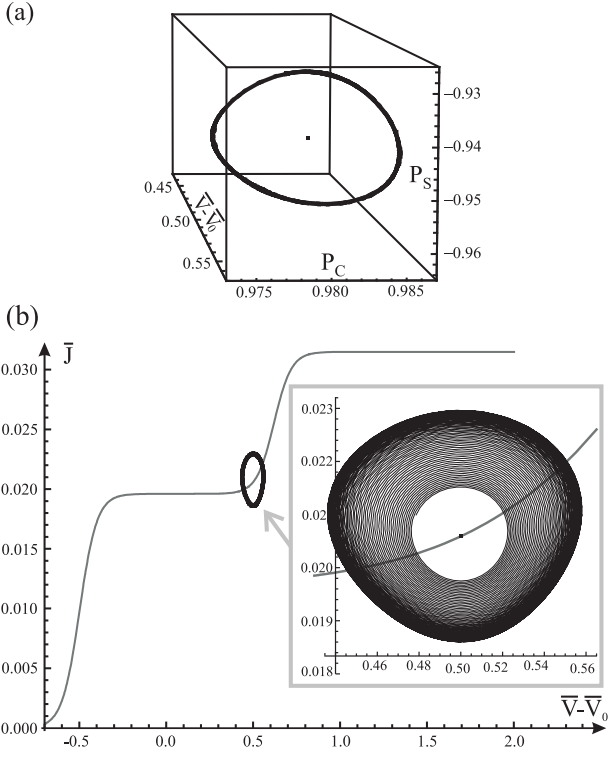}}
\caption{The limit cycle appearing as the result of the instability in the system with positive differential conductance at the stationary IVC (gray curve) in \textbf{A}-configuration, around a) the stationary points $(0.5,0.981,-0.948)$ in $(\overline{V},P_c,P_s)$ space, and b) around stationary point $(0.5,0.0206)$ in $(\overline{V},\overline{J})$ space. Used scales and voltage origin are the same as in Fig. \ref{limitcyclepic}. The choice of parameters is: $\overline{V}_b=\overline{V}_\uparrow$, $k_B T/\Delta\epsilon=0.05$, $\gamma_\uparrow=0.08$, $\gamma_\downarrow=0.02$, $\overline{\mathcal{L}}=45.0$. The critical values between which the instability takes place are $\overline{\mathcal{L}}_c=41.112$; $\overline{\omega}_c=0.5883$ and $\overline{\mathcal{L}}_c^\diamond=257.301$; $\overline{\omega}_c^\diamond=0.3272$.
The frequency of the stable oscillations $\overline{\omega}=0.5865$.}
\label{limitcycleposjmppic}
\end{figure}

The above-mentioned instability is a Hopf bifurcation, resulting in the onset of spontaneous, non-linear, periodic in time, self-excited oscillations of current $J(t)$, voltage drop $V(t)$, average charge $q(t)=e \mathcal{N} P_c(t)$ and average spin $s(t) =(1/2) \mathcal{N} P_s(t)$ in the layer of dots. In the case $(\mathcal{L}-\mathcal{L}_c)/\mathcal{L}_c \ll 1$ the frequency of the oscillations  $\approx \omega_c$. Analytical estimations and numerical calculations show  that the critical inductance increases and the oscillations fade out with an increase of the temperature, disappearing at $k_B T \sim \Delta\epsilon$. The numerical solutions of system (\ref{nonstatsys}) with $C=0$, for an N-type junction in \textbf{P}-configuration with the NDC, in circuit with inductance chosen slightly beyond the critical value, show charge/spin accumulation, as well as current, orbiting in time along the limit cycle as presented in Fig. \ref{limitcyclepic}. An example of limit cycle for \textbf{A}-configuration with positive differential conductance is shown in Fig. \ref{limitcycleposjmppic}. The results show critical inductances of the order of 1mH$/\mathcal{N}$ and critical frequencies of the order of $\Gamma_N$, i.e. 1GHz, independent of $\mathcal{N}$. Typical junction capacitance per dot $C_{1QD}\sim10^{-18}F$ introduces just small correction to critical inductance lowering it by $\sim0.2$ percent, while critical frequency remains unchanged.\\

\section{Conclusion}
We considered the effect of Coulomb blockade correlations on the spin-dependent electronic transport across a layer of quantum dots connecting a normal and a magnetic lead. It was shown that in such a system, under the voltage biasing with an inductor added in series with the junction,
an instability in the steady (time-independent) flows of charge and spin may arise. This instability develops into a new stable regime in which the average spin and charge accumulated in the dots oscillate periodically in time. The
typical frequency of the oscillations is of the order of 1GHz for realistic junction parameters.
In contrast to the standard electric instability of an RLC-circuit with a negative differential resistance, in the system under consideration spin accumulation in the dots results in an instability which occurs even in the case of an  RL-circuit with a positive differential resistance.

{\it Acknowledgement.} Financial support from the Swedish VR and SSF, the European Commission (FP7-ICT-2007-C; proj no 225955 STELE)
and the Korean WCU programme funded by MEST/NFR
(R31-2008-000-10057-0) is gratefully acknowledged.

\section{Appendix}

The rate equation

\begin{widetext}
\begin{equation}
\label{mastereq}
\frac{d}{dt}
\left(\begin{array}{c} P_0\\ P_\uparrow\\ P_\downarrow \end{array} \right)
=\sum \limits_{i=N,F}
\left(\begin{array}{ccc}
-(\Gamma_i^{\uparrow,L\rightarrow QD}+\Gamma_i^{\downarrow,L\rightarrow QD}) & \Gamma_i^{\uparrow,QD\rightarrow L} & \Gamma_i^{\downarrow,QD\rightarrow L} \\
\Gamma_i^{\uparrow,L\rightarrow QD} & -\Gamma_i^{\uparrow,QD\rightarrow L} & 0\\
\Gamma_i^{\downarrow,L\rightarrow QD} & 0 & -\Gamma_i^{\downarrow,QD\rightarrow L}
\end{array} \right)
\left(\begin{array}{c} P_0\\ P_\uparrow\\ P_\downarrow \end{array} \right).
\end{equation}
\end{widetext}

where eight $\Gamma$'s are the rates of electron tunneling between the lead ("L") and the dot ("QD"), calculated using Fermi golden rule and written in compact form as
$\Gamma_i^{\sigma,\eta}(V_i)=\frac{2\pi}{\hbar} |\tau_i|^2 \int f^\eta \left( E_{i,\sigma} (\vec{p}) +eV_i - \mu_i \right) \delta \left( E_{i,\sigma} (\vec{p}) +eV_i - \epsilon_{\sigma} \right) d\vec{p}$,
resulting in the expression

\begin{eqnarray}
\label{rates}
\Gamma_i^{\sigma,\eta} (V)=\frac{2\pi}{\hbar}|\tau_i|^2 f^\eta \left( \epsilon_\sigma + \varsigma_i eV-\mu_0 \right) \nonumber \\
\times g_i \left( \varepsilon_i + \epsilon_\sigma -\frac{1}{2}(1+\varsigma_i) \sigma I + \varsigma_i eV \right).
\end{eqnarray}

Here index $i=N,F$ denotes process between QD and N/F lead, $\sigma=\pm 1$ stands for spin $\uparrow / \downarrow$, $\varsigma_i=\mp 1$ for $i=N/F$, while $\eta=L\rightarrow QD, QD\rightarrow L$ denotes the tunneling process from the lead to the QD and vice versa respectively. $\tau_i$ is averaged, energy independent matrix element of tunneling Hamiltonian (\ref{Hamiltonians}) between QD and the lead $i$, $g_i(\varepsilon)$ is electron density of states in lead $i$ at given energy $\varepsilon$, while $f^\eta (\varepsilon)$ denotes Fermi function $f(\varepsilon)$ for $L\rightarrow QD$ process, and $1-f(\varepsilon)$ for $QD\rightarrow L$ process respectively. Since $g(V)$ is slowly varying function along the voltage interval $(V_\downarrow,V_\uparrow)$, we approximate the tunneling rates as

\begin{equation}
\label{rateaprox}
\Gamma_i^{\sigma,\eta} (V) \approx f^\eta \left( \epsilon_\sigma + \varsigma_i eV-\mu_0 \right) \Gamma_i^\sigma,
\end{equation}

where $\Gamma_i^\sigma = (2\pi / \hbar) |\tau_i|^2 g_i^\sigma$ is the voltage-independent "bare" tunneling rate, by approximating $g_i^\sigma \approx g(\varepsilon_i -\frac{1}{2}(1+\varsigma_i) \sigma I)$ for $eV$ and $\epsilon_\sigma-\mu_0 \ll \varepsilon_i,I$ ($\epsilon_\sigma-\mu_0$ can be adjusted small using the gate voltage).

\end{document}